\documentclass[
nofootinbib,
twocolumn,
showpacs,
numbers,
amsmath,
prb,
amssymb]{revtex4}
\usepackage{graphicx}
\usepackage{dcolumn}
\usepackage{bm}
\def\BiR{Bi$_2$Sr$_{2-x}R_x$CuO$_y$}
\def\La-OP{La-Bi2201-OP}
\def\Eu-OP{Eu-Bi2201-OP}
\def\Bi2212{Bi$_2$Sr$_2$CaCu$_2$O$_y$}

\def\Tc{$T_{\mbox{\scriptsize c}}$}

\def\T*{$T^*$}

\def\DSC{$\Delta_{sc0}$}

\def\Tonset{$T_{\mbox{\scriptsize onset}}$}

\begin{document}
\preprint{APS/123-QED}
\title{
%
%
Enhancement of superconducting fluctuation under the coexistence of a competing pseudogap state in \BiR\
}
\author{Y. Okada}
\affiliation{Department of Crystalline Materials Science, Nagoya University, Nagoya 464-8603, Japan}
\author{Y. Kuzuya}
\affiliation{Department of Crystalline Materials Science, Nagoya University, Nagoya 464-8603, Japan}
\author{T. Kawaguchi}
\affiliation{Department of Crystalline Materials Science, Nagoya University, Nagoya 464-8603, Japan}
\author{H. Ikuta}
\affiliation{Department of Crystalline Materials Science, Nagoya University, Nagoya 464-8603, Japan}
\begin{abstract}
The onset temperature of superconducting fluctuation \Tonset\ of \BiR\ ($R$=La and Eu)
was studied by measuring the Nernst effect.
We found that \Tonset\ has a $x$ and $R$ dependence that is 
quite different from both the pseudogap temperature \T*\ and 
the critical temperature \Tc.
Our results support the picture that 
the incoherent superconductivity, which has been observed below \Tonset,
is qualitatively different from the pseudogap phenomenon
that is characterized by \T*.
The experimentally obtained phase diagram indicates that 
the pseudogap state suppresses \Tc\ 
and enhances superconducting fluctuation 
while having only small influence on \Tonset.
\end{abstract}
\maketitle
In conventional superconductors, pairing and the phase coherence 
occur simultaneously 
when the sample is cooled through
the superconducting transition temperature \Tc.
Consequently, the superconducting gap, which corresponds to the binding energy of the paired electrons (Cooper pair), 
appears only in the superconducting state.
In high-\Tc\ cuprates, however, an energy gap in the density of states has been experimentally observed even above \Tc.
This normal-state gap is called pseudogap and 
whether it is related to fluctuation of the pairing state 
or a state that is competing with the superconducting order
has been one of the fundamental issues for high-\Tc\ cuprates for over a decade.
\cite{Emery,Chacravati,Timsk,Huffner,Norman}

Although fluctuation takes place more or less in any phase transition, 
the temperature range where it was claimed that pairing
is incoherent in cuprates is surprisingly large.
\cite{Colson}
%
It was also reported that the pseudogap has a momentum dependence 
that is similar to a $d$-wave superconducting gap.
\cite{Norman_Nature,Ding,Kanigel}
These results have been regarded as supporting
the preformed pairing picture of the pseudogap state.
However, recent studies have shown that there is no direct connection between 
the superconducting gap and the pseudogap 
in both momentum- and real-spaces 
as was demonstrated by changing carrier concentration\cite{Tanaka} 
or temperature.\cite{Lee,Kondo_PRL,Kondo_Nature,Boyer} 
A pseudogap was found also in a 
non-superconducting material,\cite{Mannella}
in favor of the view that 
the pseudogap phenomenon is not directly related to high-\Tc\ 
superconductivity.
These results suggest that the pseudogap 
has its origin in a competing state 
rather than preformed pairs.
This competing order scenario was
supported further in recent experiments 
by demonstrating the presence of a short range charge-density-wave (CDW) order.
\cite{Wise_1,Ma}
However, how this competing state suppresses high-\Tc\ superconductivity
is still not clear,
i.e., whether it affects to the pair formation
or prevents the developement of a phase coherence.

The Nernst coefficient is sensitive to superconducting fluctuation.
\cite{Xu,Wang_1,Wang_2,Li,Kontani,Ussishkin,Mello,Nernst_Patric,Nernst_Tan,Nernst_nematic,Nernst_metal,Nernst_organic} 
The Nernst signal in conventional metals is generally 
small, but it is enhanced with the growth of superconducting order when a superconducting material is cooled.
\cite{Nernst_metal,Nernst_organic,Wang_1}
Many reports agree that 
the behavior of the Nernst signal in cuprates, 
which increases continuously from a small negative value 
starting at a temperature \Tonset\ 
to a large positive value by approaching \Tc, 
can be mainly explained based on a large fluctuation of superconducting order.
\cite{Xu,Wang_1,Wang_2,Kontani,Ussishkin,Mello,Nernst_Patric,Nernst_Tan,Nernst_nematic,Li}
The data of the Nernst signal are consistent with 
the diamagnetic signal that also 
survives up to around \Tonset,
supporting the superconducting fluctuation scenario for the enhanced Nernst signal.
\cite{Wang_1,Wang_2,Li}

The reported data show that \Tonset\ lies far above \Tc\ 
but below the pseudogap temperature \T* on the phase diagram.
\cite{Wang_2,Rullier,Johannsen_Ni,Xu_Zn,Lavrov,Kudo}
However, it is still not well understood how the phenomena 
characterized by these three temperatures are related to each other.
To answer this question, we focus on the \BiR\ system.
In this system, 
\Tc\ depends on both $x$ and $R$.
\cite{Nameki, Eisaki_PRB, Fujita_PRL, Okada_PhysicaC} 
Changing $x$ alters carrier doping,
while changing $R$ varies
\Tc\ without changing carrier doping,
which provides us a unique opportunity to unveil 
how \Tonset\ changes when \Tc\ is different
while keeping the carrier concentration the same.
Combined with the results of our previous ARPES study
in which the pseudogap temperature \T*\ was determined, \cite{Okada_JPSJ}
the three temperatures \Tonset, \T*, and \Tc\ are shown to be well 
separated on the phase diagram. 
Based on these results, 
we discuss that the presence of 
three distinct temperatures is
a consequence of a coexisting and competing pseudogap state 
that brings about a large fluctuation of superconductivity.

\begin{figure}[!bp]
  \begin{center}
	\includegraphics[width=1\columnwidth,clip]{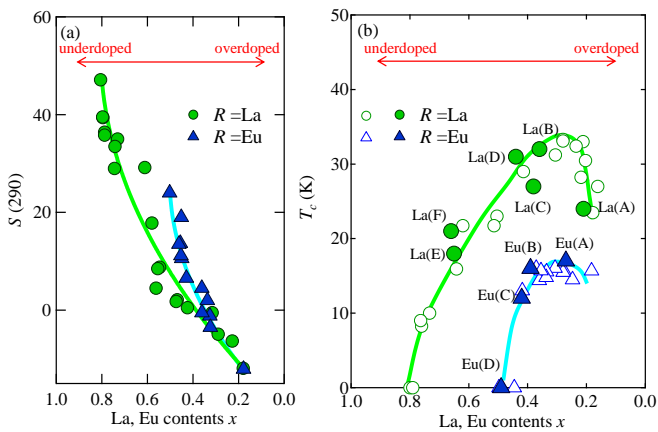}
  \end{center}
  \caption{
(Color online) 
(a) The relation between the rare earth content $x$ and the Seebeck coefficient at 290 K $S$(290).
(b) \Tc\ as a function of $x$ of the crystals used in this study (filled symbols) 
  and those in our previous study (empty symbols).\cite{Okada_JPSJ}
 }
 \label{fig:fig1.eps}
\end{figure}
\begin{figure}[!tp]
  \begin{center}
	\includegraphics[width=1\columnwidth,clip]{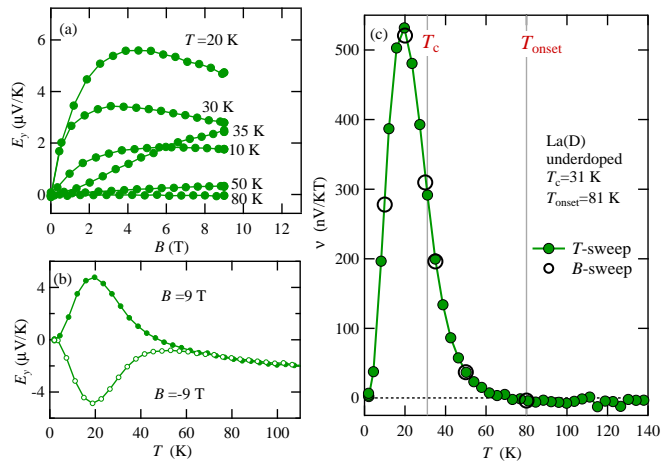}
  \end{center}
  \caption{
(Color online) The field (a) and temperature (b) dependence of the Nernst signal $E_y$ of the sample denoted by La(D) in Fig. 1(b). 
(c) A comparison of the temperature dependence of the Nernst coefficient $\nu$($T$) at 9 T determined from the field-sweep (a) and temperature-sweep (b) methods.
The dotted line is a linear extrapolation of the higher temperature data.
}
  \label{fig:fig2.eps}
\end{figure}
Single crystals of \BiR\ ($R$=La and Eu) were grown by the floating zone (FZ) method.
\cite{Okada_PhysicaC}
All the crystals used in this study were annealed with the same condition and quenched to room temperature.
\cite{Okada_PhysicaC}
Figure 1(a) shows the relation between the Seebeck coefficient at 290 K $S$(290) and $x$ of our single crystals. 
Inductively coupled plasma (ICP) spectroscopy was employed to determine $x$.
Based on the result shown in Fig. 1(a), 
the rare earth contents of the particular
samples used in the Nernst coefficient measurements were determined 
by simultaneously measuring the Seebeck coefficient.
\Tc\ of the samples used in this study are shown in Fig. 1(b) 
together with the data of the samples of our previous study.
\cite{Okada_JPSJ,notice1}

The Nernst coefficient $\nu$ 
is defined as $\nu$=$E_y$/($-$$\nabla_x$$T$)$B$
and can be obtained by measuring the electric field $E_y$, 
which is perpendicular to the magnetic field $B$ and the temperature gradient $-$$\nabla_x$$T$.
The temperature gradient was measured 
using differential type copper-constantan thermocouples.
A heating pulse was applied to generate a temperature
gradient in the sample.
The duration of the heat pulse was a few hundred seconds 
and the generated signal was measured well after the temperature 
change became small to avoid mistakenly read the voltage 
during the transient state between heater on and off.
Figures 2(a) and (b) show the field dependence of the Nernst signal $E_y$ 
at constant temperatures and the temperature dependence of $E_y$ at 9 T and $-$9 T of the sample denoted by La(D) in Fig. 1(b). 
Nernst coefficient at magnetic field $B$ was determined by calculating 
$\nu(T)=[\nu(T, B)+\nu(T, -B)]/2$ to eliminate the longitudinal signal 
and the Seebeck voltage stemming from the electrodes.
$\nu$($T$) at 9 T determined from the field and temperature dependencies are both shown in Fig. 2(c).
As shown in the figure, $\nu$($T$) obtained from the two methods coincides quite well assuring that both methods have a similar accuracy.
Therefore, we show only the Nernst coefficient 
measured by the temperature sweep mode with a constant rate (slower than 0.5 K/min) with applied fields of $|B|$=9 T in the following.
The onset temperature \Tonset\ was defined 
where the $\nu$($T$) signal deviates 
from a linear extrapolation of the higher temperature data (see Fig. 2(c)).

\begin{figure}[!ltp]
  \begin{center}
	\includegraphics[width=\columnwidth,clip]{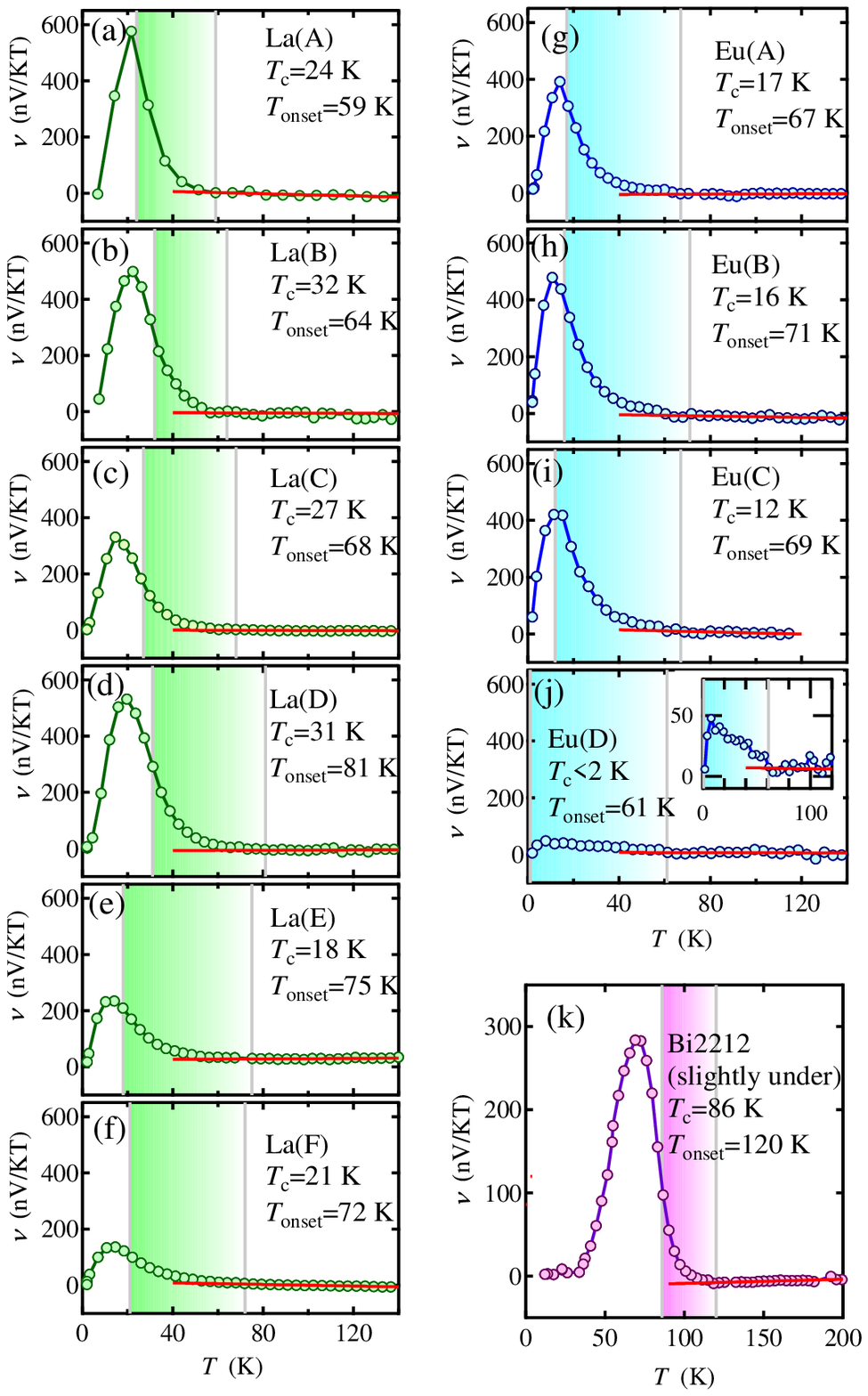}
  \end{center}
  \caption{(Color online) Temperature dependence of the Nernst coefficient $\nu$($T$) at 9 T
  of \BiR\ with $R$=La and Eu is shown in (a)-(f) and (g)-(j), respectively.
    The inset to (j) shows the data of sample Eu(D) in 
    an expanded scale.
    $\nu$($T$) of a slightly underdoped Bi2212 is shown in (k). 
    The shaded region in each figure corresponds to the temperature 
    range between 
    \Tonset\ and \Tc\ (zero field limit).
The red lines are linear extrapolations of the higher temperature data.
  }
  \label{fig:fig3.eps}
\end{figure}
Figures 3 (a)-(f) show the $\nu$($T$) curves of the \BiR\ samples with $R$=La and (g)-(j) those for $R$=Eu.
For comparison, we also measured $\nu$($T$) of a slightly underdoped 
\Bi2212\ (Bi2212, \Tc=86 K) single crystal,
and the result is shown in Fig. 3(k).
We can see from these results that the Nernst signal shows
a continuous change across \Tc\ for all samples
and the large signal below \Tc\ survives far into the normal state
(here \Tc\ refers to its zero field limit value).
This behavior is consistent with the earlier studies
that report a large superconducting fluctuation for various cuprates.
\cite{Wang_2,Rullier,Johannsen_Ni,Xu_Zn,Lavrov,Kudo}

\begin{figure}[!rbp]
  \begin{center}
	\includegraphics[width=\columnwidth,clip]{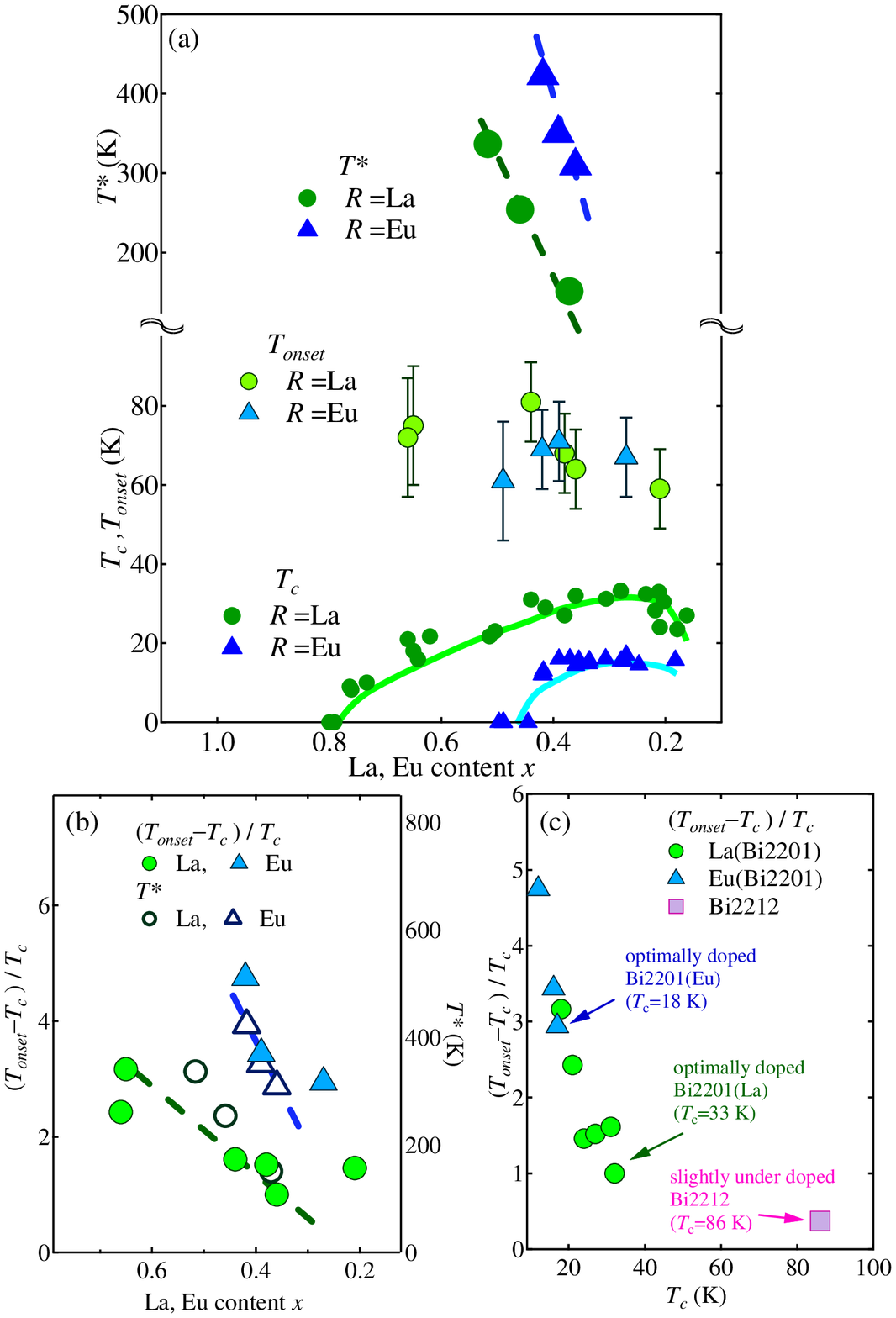}
  \end{center}
  \caption{
(Color online) (a) The phase diagram showing 
the three characteristic temperatures \T*,\cite{Okada_JPSJ} 
\Tonset, and \Tc\ of \BiR\ ($R$=La and Eu).
(b) The relation of (\Tonset$-$\Tc)/\Tc\ (left axis) and 
\T*\ (right axis) as a function of $x$.
(c) The relationship between 
(\Tonset$-$\Tc)/\Tc\ and \Tc\ of \BiR\ ($R$=La and Eu) and 
the slightly underdoped \Bi2212 (\Tc\ =86 K) sample
shown in Fig. 3 (k).
  }
  \label{fig:fig4.eps}
\end{figure}
Figure 4 (a) shows \Tonset\ of the \BiR\ samples 
together with our previously reported data of \T*\ and \Tc.
\cite{Okada_JPSJ}
As shown in Fig. 4 (a), 
the $x$ dependence of the three characteristic temperatures \Tc, \Tonset, and \T*\ are clearly different.
With decreasing carrier doping from the optimum by increasing $x$,
\Tc\ decreases and \T*\ increases
while \Tonset\ of both $R$=La and $R$=Eu samples does not change much.
At a fixed $x$,
\Tonset\ has a very similar value for $R$=La and Eu,
while \Tc\ and \T*\ depend strongly on the $R$ element.
Therefore, it is natural to think that \Tonset\ is not much affected by 
the pseudogap state that is stabilized below \T*.

However, there is an interesting connection 
between the width of the superconducting fluctuation regime and \T*.
As mentioned above, \Tonset\ of La- and Eu-doped samples are not much
different at a fixed $x$ 
despite the large difference in \Tc.
This means that the temperature range of fluctuation is wider for 
an Eu-doped sample than that of a La-doped sample when compared at 
the same $x$.
Figure 4 (b) is a plot of (\Tonset$-$\Tc)/\Tc\,
which corresponds to the width of the temperature range where
the superconducting order is fluctuating.
As shown in this figure, (\Tonset$-$\Tc)/\Tc\ and \T*\ have 
a similar $x$ and $R$ dependence, 
suggesting an intimate relation between 
the superconducting fluctuation and the pseudogap state characterized by \T*.
This result can be coherently understood
by recalling that the presence of a competing pseudogap state 
decreases the density of states that can participate in 
forming electron pairs below \Tonset.
Consequently, the overlap of the wavefunction of the paired electrons 
reduces and it becomes more difficult to form a coherent superconducting order.
As a result, \Tc\ decreases and a large superconducting fluctuation 
would be observed,
which is consistent with our result.
Here, it is known that changing the $R$ element to one with a smaller ionic radius increases disorder.\cite{Eisaki_PRB,Fujita_PRL} 
In general sence, disorder decreases superfluid density (phase stiffness), 
and hence increases the superconducting phase fluctuation. 
However, the disorder in the present case is induced to out-of-CuO$_2$ plane. 
Because this out-of-plane disorder is known as a weak scattering source,\cite{Fujita_PRL,Hashimoto_PRB} 
the naive expectation of its influence on superconductivity would be small, in contrast to the experimental results.\cite{Fujita_PRL} 
On the other hand, it has been suggested that out-of-plane disorder affects much on superconductivity via stabilizing the pseudogap state which competes with superconductivity.\cite{Okada_JPSJ}
Since the formation of the competing pseudogap state reduces the phase stiffness of superconductivity, 
we think that this is the main reason of the enhanced superconducting fluctuation observed in the present study.
Note here that \Tonset\ and (\Tonset$-$\Tc)/\Tc\ of the Bi2212 sample (Fig. 3(k)) 
were about 120 K and about 0.37 respectively,
implying that the fluctuation regime
is much narrower than Bi2201 as shown in Fig. 4(c).
This is probably related to the smaller difference in 
the magnitude of the pseudogap 
and superconducting gap of Bi2212 compared to Bi2201.
\cite{Sugimoto, Kondo_PRL}

It is worthwhile mentioning that 
the sample with the lowest carrier doping of the $R$=Eu series
showed also an increase in the Nernst signal at a temperature 
similar to \Tonset\ of the other samples 
(see the inset to Fig. 3(j)), 
although this sample did not show a superconducting transition down to 2 K.
This implies that the temperature below which superconducting fluctuation 
starts to grow is roughly the same as the optimally doped samples
even when a coherent superconducting order 
is not formed at low temperatures.
Here, it was reported for some other cuprates that \Tonset\ decreased 
in the heavily underdoped region and went to zero with \Tc.\cite{Wang_1} 
The discrepancy may stem from the difference in the doping level 
where we are focusing on in the present study since \Tc\ of Eu-doped Bi2201 
vanishes at relatively large hole doping. 
In other words, we may say from this comparison that the doping level has a strong influence on \Tonset, 
at least in the heavily underdoped region, while out-of-plane order has only small effect on \Tonset.

It is meaningful to discuss how the weak 
$R$ and $x$ dependence of \Tonset\ 
can be reconciled with the results of ARPES and STM/STS experiments.
\cite{Okada_JPSJ,Okada_LT,Okada_SNS,Okada_in_preparation,Hashimoto_PRB,Sugimoto,Machida}
Our ARPES measurement of the superconducting state 
shows that extrapolating the gap function of the node region to the antinode
gives a similar gap magnitude 
(we define this energy scale as \DSC) 
for both $R$=La and Eu.
\cite{Okada_LT,Okada_in_preparation}
According to the STM/STS experiments, 
the magnitude of the energy gap of the regions where 
relatively large coherence peaks were observed
did not depend much on the $R$ element and 
the gap size was similar to \DSC.
\cite{Sugimoto,Machida}
These weak $R$ dependence of \DSC\
both in momentum- and real-spaces are probably 
related to the weak $R$ dependence of the onset
temperature of pairing \Tonset.
Interestingly, the mean field superconducting transition temperature calculated 
from \DSC\ is comparable to \Tonset.
\cite{Okada_in_preparation}
We think that this agreement suggests that 
\DSC\ corresponds to the energy of pairing 
that starts to form at \Tonset\ when the temperature is lowered.
\cite{Sugimoto,Machida}
On the other hand, it has been reported that 
both the pseudogap at the antinode\cite{Hashimoto_PRB,Okada_LT,Okada_SNS,Okada_in_preparation} 
and the spatially averaged energy gap\cite{Sugimoto}
increase with decreasing the ionic radius of the $R$ element.
Therefore, it is natural to think that the gap at the antinode 
is related to \T*\ and brought about 
the deviation from the mean field picture 
by enhancing the superconducting fluctuation.

In summary, we studied the onset temperature of 
superconducting fluctuation \Tonset\ 
by Nernst effect measurements of \BiR.
The experimentally obtained phase diagram 
shows clearly that 
the three characteristic temperatures \T*, \Tonset, and \Tc\ 
are different irrespectively to the carrier content 
$x$ and the $R$ element. 
The results indicate that 
the pseudogap state suppresses superconductivity 
while having little influence on the onset temperature of pairing,
which in turn causes a large enhancement of superconducting fluctuation.

This work is partially supported by the Japan Society for the Promotion of Science 
under the Grant-in-Aid for JSPS Fellows 
of Ministry of Education, Culture, Sports, Science and Technology.



\begin{thebibliography}{99}
    \bibitem{Emery}
        V. J. Emery, and S. A. Kivelson,
        Nature {\bf 374}, 434 (1995).
    \bibitem{Chacravati}
        S. Chakravarty, R. B. Laughlin, D. K. Morr, and Chetan Nayak,
        Phys. Rev. B {\bf 63}, 094503 (2001).
    \bibitem{Timsk}
        T. Timusk, and B. Statt, Rep. Prog. Phys. 
        {\bf 62}, 61 (1999).
    \bibitem{Huffner}
        S. Huffner, M. A. Hossain, A. Damascelli, and G. A. Sawatzky, 
        Rep. Prog. Phys. {\bf 71}, 062501 (2008).
    \bibitem{Norman}
        M. R. Norman, D. Pines, and C. Kallin, 
        Advances in Physics, {\bf 54}, 715 (2005). 
    \bibitem{Colson}
        J. Corson, R. Mallozzi, J. Orenstein, 
        J. N. Eckstein, and I. Bozovic, 
        Nature {\bf 398}, 221 (1999).
    \bibitem{Norman_Nature}
        M. R. Norman, H. Ding, M. Randeria, J. C. Campuzano, 
        T. Yokoya, T. Takeuchi, T. Takahashi, T. Mochiku, 
        K. Kadowaki, P. Guptasarma, and D. G. Hinks, 
        Nature, {\bf 392}, 157 (1998).
    \bibitem{Ding}
        H. Ding, T. Yokoya, J. C. Campuzano, T. Takahashi, 
        M. Randeria, M. R. Norman, T. Mochiku, K. Kadowaki, and J. Giapintzakis,
        Nature {\bf 382}, 51 (1996).
    \bibitem{Kanigel}
        A. Kanigel, M. R. Norman, M. Randeria, U. Chatterjee, S. Souma, A. Kaminski, 
        H. M. Fretwell, S. Rosenkranz, M. Shi, T. Sato, T. Takahashi, Z. Z. Li, H. Raffy, K. Kadowaki,
        D. Hinks, L. Ozyuzer, and J. C. Campuzano, 
        Nature Phys. {\bf 2}, 447 (2006).
    \bibitem{Tanaka}
        K. Tanaka, W. S. Lee, D. H. Lu, A. Fujimori, T. Fujii, 
        Risdiana, I. Terasaki, D. J. Scalapino, T.   P. Devereaux, 
        Z. Hussain, and Z.-X. Shen, 
        Science {\bf 314}, 1910 (2006).
    \bibitem{Kondo_PRL}
        T. Kondo, T. Takeuchi, A. Kaminski, S. Tsuda, and S. Shin, 
        Phys. Rev. Lett. {\bf 98}, 267004 (2007).
    \bibitem{Lee}
        W. S. Lee, I. M. Vishik, K. Tanaka, D. H. Lu, T. Sasagawa, N. Nagaosa, 
        T. P. Devereaux, Z. Hussain, and Z.-X. Shen, 
        Nature {\bf 450}, 81 (2008).
    \bibitem{Kondo_Nature}
        T. Kondo, R. Khasanov, T. Takeuchi, J. Schmalian, and A. Kaminski,
        Nature {\bf 457}, 296 (2008).
    \bibitem{Boyer}
        M. C. Boyer, W. D. Wise, K. Chatterjee, M. Yi, T. Kondo, 
        T. Takeuchi, H. Ikuta, and E. W. Hudson, 
        Nat. Phys. {\bf 3}, 802 (2007).
    \bibitem{Mannella}     
        N. Mannella, W. Yang, X. J. Zhou, H. Zheng, J. F. Mitchell, J. Zaanen, 
        T. P. Devereaux, N. Nagaosa, Z. Hussain, Z.-X. Shen, 
        Nature {\bf 438}, 474 (2005).
    \bibitem{Wise_1}
        W. D. Wise, M. C. Boyer, K. Chatterjee, T. Kondo, T. Takeuchi, 
        H. Ikuta, Y. Y. Wang, and E. W. Hudson, 
        Nat. Phys. {\bf 4}, 696 (2008).
    \bibitem{Ma}
        J.-H. Ma, Z.-H. Pan, F. C. Niestemski, M. Neupane, Y.-M. Xu, P. Richard, 
        K. Nakayama, T. Sato, T. Takahashi, H.-Q. Luo, L. Fang, H.-H. Wen, Ziqiang Wang, 
        H. Ding, and V. Madhavan,
        Phys. Rev. Lett. {\bf 101}, 207002 (2008).
    \bibitem{Xu}
        Z. A. Xu, N. P. Ong, Y. Wang, T. Kakeshita, and S. Uchida, 
        Nature {\bf 406}, 486 (2000).
    \bibitem{Wang_2}
        Y. Wang, L. Li, M. J. Naughton, G. D. Gu, S. Uchida, and N. P. Ong,
        Phys. Rev. Lett. {\bf 95}, 247002 (2005).
    \bibitem{Wang_1}
        Y. Wang, L. Li, and N. P. Ong, 
        Phys. Rev. B {\bf 73}, 024510 (2006).
    \bibitem{Nernst_metal}
        A. Pourret, H. Aubin, J. Lesueur, 
        C. A. Marrache-Kikuchi, L. Berg\'e, L. Dumoulin, and K. Behnia, 
        Nature Phys. {\bf 2}, 683 (2006).
    \bibitem{Nernst_organic}
        M. S. Nam, A. Ardavan, S. J. Blundell, and J. A. Schlueter,
        Nature {\bf 449}, 584 (2007).
    \bibitem{Kontani}
        H. Kontani, 
        Phys. Rev. Lett. {\bf 89}, 237003 (2002).
    \bibitem{Ussishkin}
        I. Ussishkin, S. L. Sondhi, and D. A. Huse, 
        Phys. Rev. Lett. {\bf 89}, 287001 (2002).
    \bibitem{Nernst_Tan}
        S. Tan, K. Levin, 
        Phys. Rev. B {\bf 69}, 064510 (2004).
    \bibitem{Nernst_Patric}
        C. Honerkamp and P. A. Lee, 
        Phys. Rev. Lett. {\bf 92}, 177002 (2004).
    \bibitem{Mello}
        E. V. L. Mello, and D. N. Dias, 
        J. Phys. Condens. Matter {\bf 19}, 086218 (2007).
    \bibitem{Nernst_nematic}
        R. Daou, J. Chang, David LeBoeuf, Olivier Cyr-Choiniere, 
        Francis Laliberte, Nicolas Doiron-Leyraud, B. J. Ramshaw, Ruixing Liang, 
        D. A. Bonn, W. N. Hardy, Louis Taillefer, 
        Nature {\bf 463}, 08716 (2010).
    \bibitem{Li}
        Lu Li, Yayu Wang, Seiki Komiya, Shimpei Ono, Yoichi Ando, G. D. Gu, and N. P. Ong, 
        arXiv:0906.1823                
    \bibitem{Rullier}
        F. Rullier-Albenque, R. Tourbot, H. Alloul, P. Lejay, 
        D. Colson, and A. Forget,
        Phys. Rev. Lett. {\bf 96}, 067002 (2006).
    \bibitem{Johannsen_Ni}
        N. Johannsen, Th. Wolf, A. V. Sologubenko, T. Lorenz, 
        A. Freimuth, and J. A. Mydosh,
        Phys. Rev. B. {\bf 76}, 020512(R) (2007).
    \bibitem{Xu_Zn}
        Z. A. Xu, J. Q. Shen, S. R. Zhao, Y. J. Zhang, and C. K. Ong,
        Phys. Rev. B. {\bf 72}, 144527 (2005).
    \bibitem{Lavrov}   
        A. N. Lavrov, Y. Ando and S. Ono,
        Europhys. Lett. {\bf 57}, 267 (2002).
    \bibitem{Kudo}
        K. Kudo, Y. Miyoshi, T. Sasaki, T. Nishizaki, N. Kobayashi,
        J. Phys. Soc. Jpn. {\bf 75}, 124710 (2006).    
    \bibitem{Nameki}
        H. Nameki, M. Kikuchi, and Y. Syono, 
        Physica C {\bf 234}, 255 (1994).
    \bibitem{Eisaki_PRB}
        H. Eisaki, N. Kaneko, D. L. Feng, A. Damascelli, 
        P. K. Mang, K. M. Shen, Z.-X. Shen, and M. Greven, 
        Phys. Rev. B {\bf 69}, 064512 (2004).
    \bibitem{Fujita_PRL}
        K. Fujita, T. Noda, K. M. Kojima, 
        H. Eisaki, and S. Uchida, 
        Phys. Rev. Lett. {\bf 95}, 097006 (2005).
    \bibitem{Okada_PhysicaC}
        Y. Okada and H. Ikuta, 
        Physica C {\bf 445-448}, 84 (2006).
    \bibitem{Okada_JPSJ}
        Y. Okada, T. Takeuchi, T. Baba, S. Shin, and H. Ikuta,
        J. Phys. Soc. Jpn. {\bf 77}, 074714 (2008).
    \bibitem{notice1}
        The carrier concentration depends not only on $x$ 
        but also on the oxygen content $y$ in this material.
        %
        However, since we annealed all the samples with the same condition, 
        $x$ can be used as a qualitative measure of the carrier concentration 
        within the purpose of this study.
    \bibitem{Sugimoto}
        A. Sugimoto, S. Kashiwaya, H. Eisaki, H. Kashiwaya, 
        H. Tsuchiura, Y. Tanaka, K. Fujita, and S. Uchida, 
        Phys. Rev. B {\bf 74}, 094503 (2006).

    \bibitem{Okada_LT}
        Y. Okada, T. Takeuchi, M. Ohkawa, A. Shimoyamada, 
        K. Ishizaka, T. Kiss, S. Shin, and H. Ikuta, 
        arXiv:0807.2165 
    \bibitem{Okada_SNS}
        Y. Okada, T. Takeuchi, A. Shimoyamada, 
        S. Shin, and H. Ikuta, 
        J. Phys. Chem. Solids {\bf 69}, 2989 (2008).
    \bibitem{Hashimoto_PRB}
        M. Hashimoto, T. Yoshida, A. Fujimori, D.H. Lu, 
        Z.-X. Shen, M. Kubota, K. Ono, 
        M. Ishikado, K. Fujita, and S. Uchida, 
        Phys. Rev. B {\bf 79}, 144517 (2009).
    \bibitem{Machida}
        T. Machida, Y. Kamijo, K. Harada, T. Noguchi,
        R. Saito, T. Kato, and H. Sakata, 
        J. Phys. Soc. Jpn. {\bf 75}, 083708 (2006).
    \bibitem{Okada_in_preparation}
        Y. Okada $et$ $al.$ (in preparation)
\end{thebibliography}
\end{document}